\documentclass[prl,showpacs,twocolumn]{revtex4}
\usepackage{slashbox}
\usepackage{amsmath,graphicx,amsfonts,bm,amssymb}
\usepackage{times}

\begin{document}

\title{Detecting non-Abelian statistics of Majorana fermions in quantum nanowire networks}

\author{Zheng-Yuan Xue}

\affiliation{Laboratory of Quantum Information Technology, and
School of Physics and Telecommunication Engineering, South China
Normal University, Guangzhou 510006,  China}


\begin{abstract}
We propose a scheme in semiconducting quantum nanowires structure to
demonstrate  the non-Abelian statistics for Majorana fermions in
terms of braid group. The Majorana fermions are localized at the
endpoints of semiconducting wires, which are deposited on an
\emph{s}-wave superconductor.  The non-Abelian nature of Majorana
fermion is manifested by the fact that the output of the different
applied orders of two operations, constructed by the braid group
elements, are different. In particular, the difference can be
unambiguously imprinted on the quantum states of a superconducting
flux qubit.
\end{abstract}


\date{\today}

\maketitle

Quantum statistics is a fundamental concept in physics which
distinguishes fermions from bosons in three dimensions. For
quasi-particles live in two dimensions, they may have two different
classes of exotic statistics: Abelian or non-Abelian statistics.
When one particle is exchanged in a counterclockwise manner with the
nearest particle, the relation between the initial wave function
$\psi(\textbf{r}_i, \textbf{r}_j)$ with the final wave function
$\psi'(\textbf{r}_i, \textbf{r}_j)$ is given by $\psi'(\textbf{r}_i,
\textbf{r}_j) = M_{ij}\psi(\textbf{r}_i, \textbf{r}_j)$, where
$\textbf{r}_i$ is the position of the particle $i$. The particles
are called Abelian anyons \cite{ab} with the statistics angle
$\theta$ if $M_{ij} = \exp(\text{i}\theta)$ where $\theta$ is not
zero or $\pi$ ($\theta= 0$ and $\pi$ correspond to bosons and
fermions, respectively). Furthermore, $M_{ij}$ can be the elements
of a braid group, where two elements $M_{ij}$ and $M_{i'j'}$ may
even be non-commutative, and quasi-particles with such features are
called non-Abelian anyons \cite{nonab}.

Potential host systems for the exotic non-Abelian statistics
including the $\nu=5/2$ quantum Hall state \cite{qhs1,qhs2}, chiral
\emph{p}-wave superconductors \cite{p}, topological
insulator-superconductor  \cite{ti} and semiconductor-superconductor
structures \cite{semi1,semi2,semi3}. With the potential applications
in topological quantum computation, Majorana fermions (MF) with
non-Abelian statistics have attracted strong renewed interests. MF
are a kind of self-conjugate quasi-particles induced from a vortex
excitation in $p_x+ip_y$ superconductor. However, due to the
instability of the $p$-wave superconducting states, its
implementation still remains an experimental challenge. Therefore,
setups with $s$-wave superconductor proximity effect
\cite{ti,semi1,semi2,semi3}, which is more stable, is more
preferred. In principle, they should allow robust topological
superconducting phase without unrealistic experimental conditions.
More recently, it is recognized that topologically protected states
may be most easily engineered in 1D semiconducting wires deposited
on an \emph{s}-wave superconductor \cite{1d1,1d2,1d3}: the endpoints
of such wires support localized zero-energy MF. This setup provides
the first realistic experimental setting for Kitaev's 1D topological
superconducting state \cite{kitaev}: MF can be created, transported,
fused and braided by applying locally tunable gates to the wire.
However, detecting MF in such setup is a challenge task because they
hold neutral charge. Recently, it has been shown that combine two MF
into a single Dirac fermion allows the neutral quasi-particles to be
probed with charge transport \cite{detect1,detect2,detect3,detect}.
But, they suffer from the same impediment: Abrikosov vortices are so
massive objects that behave classically. Fortunately, alternatives
using Josephson vortexes are proposed by introducing a controllable
superconducting flux qubit \cite{detect4,detect5,detect6}, which
enable one to detect unambiguously the states of MF in this 1D
scenario.

Recently, Zhu \emph{et al.} \cite{zhu}  proposed a scheme to
directly test the quantum statistics of the braid group for MF in
cold atoms scenario. Confirming this aspect of MF is not only of
significant important in its own, but also the crucial and first
step towards the realization of topological quantum computation.
However, simulate exotic statistics of MF in a macroscopic material
is another story. Here, we propose such an alternative scheme to
detect the non-Abelian statistics of the MF in terms of braid group
in semiconducting quantum nanowires structure. The non-Abelian
nature is manifested by the fact that the output of the different
applied orders of two operations are different, which is different
form previous schemes based on charge transport
\cite{detect1,detect2,detect3,detect,detect4}. The operations are
constructed by the braid group elements of MF. Furthermore, the two
different final states can be distinguished by measuring the states
of MF by the superconducting flux qubit.

\begin{figure}[tbp]\centering
\includegraphics[width=.7\columnwidth]{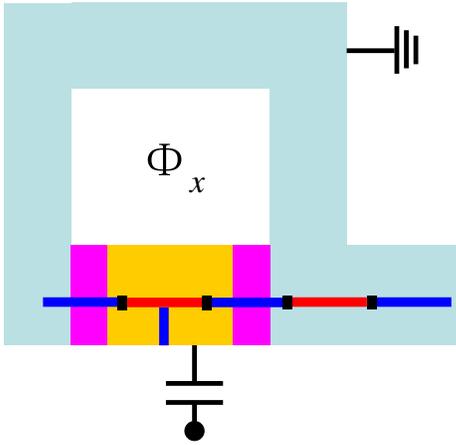}
\caption{(Color online) Superconducting flux qubit with two
Josephson junctions (pink)  and an enclosing magnetic flux $\Phi_x$.
MF (black) are induced at the interface between a topologically
trivial (blue) and a topologically nontrivial (red) section of an
quantum nanowire. Gate electrodes (not shown) can be used to move
the MF along the wire.} \label{fig1}
\end{figure}

The setup we consider is shown in Fig. \ref{fig1}, which is  a
spin-orbit coupled semiconducting wire deposited on an \emph{s}-wave
superconductor. Applying a magnetic field perpendicular to the
superconductor surface, the Hamiltonian describing such a wire is
\cite{1d1}
\begin{eqnarray}
H&=& \int \bigg{[} \psi^\dagger_x
\bigg(-\frac{\hbar^2\partial_x^2}{2m}-\mu - i \hbar u \bold{\hat
e}\cdot \bm{\sigma}\partial_x +V_B \sigma_z\bigg)\psi_x
  \nonumber \\
  && + (|\Delta| e^{i\varphi} \psi_{\downarrow x} \psi_{\uparrow x} +
  h.c.)\bigg{]}dx,
  \label{h}
\end{eqnarray}
where $\psi_{\alpha x}$ corresponds to electrons with spin $\alpha$,
effective mass $m$, and chemical potential $\mu$; the third term
denotes spin-orbit coupling with $u$ the strength, and $\bm{\sigma}
= (\sigma_x,\sigma_y,\sigma_z)$ is the vector of Pauli matrices; the
fourth term represents the energy shift due to the magnetic field;
and the terms in the second line are the spin-singlet pairing from
the $s$-wave superconductor via proximity effect.

In the setup, the magnetic field is weak, and the superconductor of
the flux qubit is the conventional $s$-wave superconductor. The
interplay of Zeeman effect, spin-orbit coupling, and the proximity
to an s-wave superconductor drive the wire into a chiral $p$-wave
superconducting state \cite{1d1,1d2,1d3}, providing that the wire is
long compared to the superconducting coherence length ($\xi\simeq
40$ nm for the superconducting substrate being Nb). Specifically,
when $|\Delta| < |V_B|$, $\mu$ lies inside of the zero-mode energy
gap, the wire is in the Kitaev's topological phase \cite{kitaev},
which supports Majorana modes; otherwise the system is an ordinary
superconductor (topological trivial). The zero-mode excitation gap
and $\mu$ dependence is \cite{1d3} $E_0 =
\left||V_B|-\sqrt{|\Delta|^2 + \mu^2}\right|$. For $|\mu|<\mu_c =
\sqrt{V_B^2 -|\Delta|^2}$ the topological phase with end MFs emerge,
or a topologically trivial phase. Thus, applying a gate voltage
uniformly allows one to create or remove the MF.  To avoid gap
closure, A "keyboard" of local tunable gate electrodes to the wire
\cite{1d3} is used to control whether a region of the wire is
topological or not. For InAs quantum narowire, assuming $|V_B| \sim
2|\Delta|$ and $\hbar u \sim 0.1$eV{\AA}, the gap for a 0.1$\mu$m
wide gate is of order 1K \cite{1d3}. It is note that heavy-element
wires and/or narrower gates could generate even larger gap.
Therefore, MF are induced at the interface between topologically
trivial and nontrivial sections of the quantum nanowire.

For a pair of MF, they can be combined to form a complex fermionic
states $c=\gamma_2+i\gamma_1$, which can be occupied $|1\rangle$ or
unoccupied $|0\rangle$, differ in fermion parity, and therefore
2-fold degenerated. A winding of one MF around another is associated
with a unitary transformation in the subspace of degenerated ground
states. For \emph{2N} MF, such unitary transformations form a set
named braid group, which is generated by elementary interchanges of
neighboring MF \cite{ivanov}.

For our purpose to verify the no-Abelian nature of MF, four MF is
enough. They combine into two complex fermions $c_1$ and $c_2$ and
the ground state has degeneracy four, but the Hamiltonian (\ref{h})
conserves parity of the fermion number, and the even-number subspace
is decoupled from the odd-number subspace. At low temperature
(significantly below the zero-mode gap), the initial state is
typically a vacuum state $|0\rangle$, and in the even-number
subspace $\{|0\rangle|0\rangle, |1\rangle|1\rangle\}$, only two of
the three generators are independent:
\begin{eqnarray}
\tau_1=\tau_3={1 \over \sqrt{2}}\left(\begin{array}{cc}
                      1-i & 0 \\
                      0 & 1+i
                    \end{array}
\right), \quad \tau_2={1 \over \sqrt{2}}\left(\begin{array}{cc}
                      1 & -i \\
                    -i & 1
                    \end{array}
\right).
\end{eqnarray}
From these, two composite braiding operations \cite{zhu}
\begin{eqnarray} \label{operation}
A=\tau_1\tau_2\tau_1^{-1}, \quad B=\tau_1^{-1}\tau_2\tau_1^{-1}
\end{eqnarray}
can be constructed with the property $AB=i\sigma_z$,
$BA=-i\sigma_x$, i.e., $AB\neq BA$, manifested the non-Abelian
nature of MF, where $\tau_i$ and $\tau_i^{-1}$ denote
cunter-clockwise and clockwise interchange of MF $i$ and $i+1$,
respectively. Specifically, with the initial state $|0\rangle
|0\rangle $, $AB$ and $BA$ yield two orthogonal output states
$|0\rangle |0\rangle $ and $|1\rangle |1\rangle $, respectively. We
on purposely chose the two outputs to be orthogonal with each other
so that the difference between the two can be detected
unambiguously.

However, to implement braiding of MF in a 1D structure is
impossible. Therefore, in order to exchange the MF for braiding
operations, one can use a second wire to form a T-junction or more
efficiently with the "railroad track" geometry \cite{1d3}. The two
red sections  in Fig. \ref{fig1} is tune into the topological phase
while the blue sections are not. In this way, we induce two pairs of
MF with the left pair (labeled as 1 and 2) residents in the
superconducting island of the flux qubit; the right pair of MF is
labeled as 3 and 4. Another wire, perpendicular to the wire holds
MF, is introduced to enable interchange of neighboring MF. In our
setup, to demonstrate the non-Abelian nature of MF, i.e., to
implement braiding operations in Eq. (\ref{operation}), we need to
counterclockwise interchange of MF 1, 2 and 2, 3 for $\tau_1$ and
$\tau_2$, respectively. Counterclockwise interchange of MF can be
achieved along the line as proposed in Ref. \cite{1d3}. For example,
$\tau_1$ can be implemented as following:  transports MF 1 downward
by driving the vertical wire into a topological phase; transports MF
2 leftward in a similar fashion; and finally transports MF 1 up and
to the right. Similarly, $\tau_1^{-1}$ can be implemented as
following: transports MF 2 downward, transports MF 1 rightward  and
finally transports MF 2 up and to the left.

We now turn to detect the two orthogonal output states
$|0\rangle_L|0\rangle_R$ and $|1\rangle_L|1\rangle_R$ . As the
output states of the two pairs of MF are the same, detecting one of
them can then fulfil the purpose of distinguishing the output
states. The two states $|0\rangle$ and $|1\rangle$ are distinguished
by the parity of the number $n_p$ of particles in the island.
Therefore, they can be distinguished by $n_p$. As
$|0\rangle_L|0\rangle_R$ and $|1\rangle_L|1\rangle_R$ have the same
parity, we need to remove one pair of MF out of the measurement
circle, otherwise they will also contribute to the measurement
results and make the detection impossible. Without loss of
generality, we choose to detect the left pair of MF (MF 1 and 2)
while move MF 3 and 4 along the wire out of the flux qubit circuit,
as shown in Fig. \ref{fig1}, by local tunable gates. Note that the
wire is not interrupted by the junctions providing that the
junctions' thickness is much smaller than $\xi$.

\begin{figure}[tbp]\centering
\includegraphics[width=.7\columnwidth]{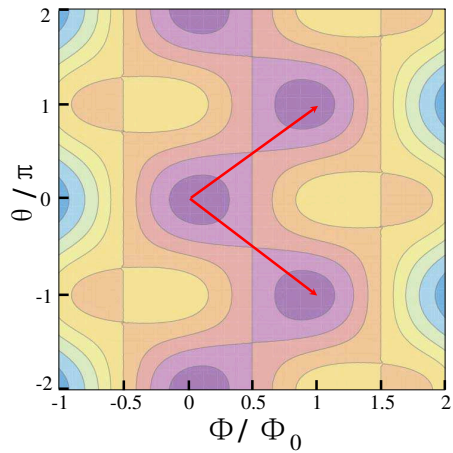}
\caption{(Color online) Potential of the symmetric superconducting
flux qubit with $\Phi_x=\Phi_0/2$ and $E_J=0.507 \Phi_0^2/(2L)$. The
energy minima are connected by two tunneling paths indicated by the
red arrows.} \label{u}
\end{figure}

To measure the parity of $n_p$, we make use of the suppression of
macroscopic quantum tunneling by the Aharonov-Casher effect
\cite{ac}: a vortex encircling a superconducting island picks up a
phase increment $\phi=\pi q/e$ determined by the total charge $q$
coupled capacitively to the superconductor. For our case, the two
Josephson junctions  have the same Josephson coupling energy $E_J$.
The flux $\Phi$ in the qubit is related to the phase differences
across the junctions, $\varphi_1$ and $\varphi_2$, by
$2\pi\Phi/\Phi_0=\varphi_1+\varphi_2$; the phase
$\theta=(\varphi_1-\varphi_2)/2$ is conjugate to the number of
excess Cooper pairs of the superconducting island as $[\theta,
n]=i$. The potential energy is plotted in Fig. \ref{u} for the
external magnetic $\Phi_x=\Phi_0/2$ and $E_J=0.507 \Phi_0^2/(2L)$
with $L$ being the inductance of the flux qubit, which is strictly
periodic in the $\theta$ direction. Therefore, neighboring minima
are always separated by $\delta\theta=\pm\pi$ (as indicated by red
arrows in Fig. \ref{u}), i.e., the energy minima are connected by
two tunneling paths with same amplitude, which amounts to the
circulation of a Josephson vortex around the superconducting island.
The interference produces an oscillatory tunnel splitting of the two
levels of the flux qubit
\begin{equation}
\Delta  = \Delta_{0}  \cos\left(\phi\over 2\right),
\end{equation}
where $ \Delta_{0}$ is the tunnel splitting associated with one
path. Therefore, if $q$ is an odd (even) multiple of the electron
charge $e$, the two tunneling paths interfere destructively
(constructive), so the tunnel splitting   is minimum (maximal).

As we only need to distinguish maximal from minimal tunnel
splitting, the flux qubit does not need to have a large quality
factor.  In addition, $ \Delta_{0}\simeq 100 \mu{\rm eV}\simeq 1$K
for parameters in typical experiments of flux qubits \cite{ac},
which should be readily observable by microwave absorption. To make
sure the total charge is solely comes from $|1\rangle_L$, one would
first calibrate the charge on the gate capacitor to zero, by
maximizing the tunnel splitting in the absence of vortices in the
island. Meanwhile, the read-out is nondestructive, which is
necessary for the realization of a two-qubit CNOT gate
\cite{detect4}. Moreover, the read out is insensitive to sub-gap
excitations in the superconductor (they do not change the fermion
parity).

In summary, we have proposed a scheme in
semiconductor-superconductor structure to demonstrate the
non-Abelian statistics for MF in terms of braid group. The
non-Abelian nature is manifested by the fact that the output of the
different applied orders of two operations, constructed by the braid
group elements, are different. Meanwhile, the different final states
of MF can be unambiguously detected by a superconducting flux qubit.

\bigskip

This work was supported by the NSFC (No. 11004065), the NSF of
Guangdong Province (No. 10451063101006312), and the Startup
Foundation  of SCNU (No. S53005).

\end{document}